\documentclass[runningheads]{llncs}
\usepackage[T1]{fontenc}
\usepackage{graphicx}
\usepackage{amsmath, amssymb, amsfonts, bm, mathtools}
\usepackage{hyperref}
\usepackage{fontawesome5}
\usepackage[table]{xcolor}  
\usepackage{soul, color}
\usepackage{mdframed}
\usepackage{graphicx}
\usepackage{subcaption}
\usepackage{caption}
\usepackage{amsmath}
\usepackage{mathabx}
\usepackage{mdframed}
\usepackage{listings}
\usepackage[utf8]{inputenc}
\usepackage{xcolor}

\begin{document}
\title{Knowledge-to-Data: LLM-Driven Synthesis of Structured Network Traffic for Testbed-Free IDS Evaluation}

%
%
\author{Konstantinos E. Kampourakis\inst{1}\orcidID{0009-0000-8883-0735} \and
Vyron Kampourakis\inst{1}\orcidID{0000-0003-4492-5104}\faIcon{envelope} \and
Efstratios Chatzoglou\inst{2}\orcidID{0000-0001-6507-5052} \and
Georgios Kambourakis\inst{2}\orcidID{0000-0001-6348-5031}
\and Stefanos Gritzalis \inst{3}\orcidID{0000-0002-8037-2191}}
\authorrunning{Konstantinos E. Kampourakis et al.}
%
\institute{Norwegian University of Science
and Technology, 2802 Gjøvik, Norway \email{\{konstantinos.kampourakis, vyron.kampourakis\}@ntnu.no} \and University of the Aegean, 83200 Karlovasi, Greece \email{\{efchatzoglou, gkamb\}@aegean.gr} \and University of Piraeus, Greece \email {sgritz@unipi.gr}}
\maketitle              
\begin{abstract}

Realistic, large-scale, and well-labeled cybersecurity datasets are essential for training and evaluating Intrusion Detection Systems (IDS). However, they remain difficult to obtain due to privacy constraints, data sensitivity, and the cost of building controlled collection environments such as testbeds and cyber ranges. This paper investigates whether Large Language Models (LLMs) can operate as controlled knowledge-to-data engines for generating structured synthetic network traffic datasets suitable for IDS research. We propose a methodology that combines protocol documentation, attack semantics, and explicit statistical rules to condition LLMs without fine-tuning or access to raw samples. Using the AWID3 IEEE~802.11 benchmark as a demanding case study, we generate labeled datasets with four state-of-the-art LLMs and assess fidelity through a multi-level validation framework including global similarity metrics, per-feature distribution testing, structural comparison, and cross-domain classification. Results show that, under explicit constraints, LLM-generated datasets can closely approximate the statistical and structural characteristics of real network traffic, enabling gradient-boosting classifiers to achieve F1-scores up to 0.956 when evaluated on real samples. Overall, the findings suggest that constrained LLM-driven generation can facilitate on-demand IDS experimentation, providing a testbed-free, privacy-preserving alternative that overcomes the traditional bottlenecks of physical traffic collection and manual labeling.

\end{abstract}

\section{Introduction}
\label{S:intro}

Almost twenty years after Clive Humby~\cite{humby2006data} coined the phrase ``data is the new oil'', data has become an indispensable asset driving technological innovation and progress. The disruptive advances observed in recent years have heavily relied on the availability of vast, diverse, and high-quality training data; most notably, the emergence of Generative Artificial Intelligence (GenAI) and the rapid evolution of Large Language Models (LLMs). The same centrality of data applies to the domain of cybersecurity, where well-established datasets are critical for designing, training, and evaluating Intrusion Detection Systems (IDS) and threat monitoring schemes. In particular, machine learning–based security solutions critically rely on representative network traffic data to accurately model normal behavior and distinguish it from malicious activity.

However, obtaining realistic, large-scale, and well-labeled cybersecurity datasets remains a persistent challenge due to several reasons, including privacy concerns, data sensitivity, and the operational cost of setting up controlled ecosystems, such as testbeds~\cite{AWID3,gotham2024}, digital twins~\cite{swat2016,kampDT2025}, or cyber ranges~\cite{kampCR2025}. To this end, synthetic data generation has emerged as a promising alternative for addressing these limitations, supporting experimentation beyond what is feasible with real-world traces alone. Although traditional synthetic data generation approaches, such as simulators, probabilistic models, or traffic replay, have shown great utility in controlled settings, they frequently struggle to capture complex inter-feature dependencies, protocol constraints, and high-level behavioral semantics simultaneously.

In this context, recent advances in LLMs open a new avenue for structured synthetic data generation~\cite{nadas2025}. Specifically, beyond their success in natural language processing, modern LLMs exhibit strong capabilities in knowledge integration, constraint reasoning, and the generation of structured outputs. These properties suggest that LLMs may function as knowledge-to-data engines, capable of transforming textual specifications, statistical constraints, and domain expertise into realistic, high-dimensional datasets that reflect both low-level feature distributions and high-level behavioral patterns. Unlike traditional synthetic data generation approaches, LLMs can potentially encode protocol semantics, temporal dependencies, and cross-feature relationships directly from descriptive inputs. This observation motivates this study to answer a fundamental research question (RQ), as seen below.

\begin{mdframed}[
  backgroundcolor=gray!10,
  linecolor=black,
  linewidth=0.4pt,
  nobreak=true
]
\textbf{\textit{RQ:}} To what extent can LLM-generated synthetic data faithfully preserve the statistical, structural, and semantic characteristics of real network traffic, thereby serving as a reliable substitute for real-world datasets in intrusion detection research?
\end{mdframed}

\noindent \textbf{\textit{Contribution:}} We propose a controlled methodology for LLM-driven synthetic labeled dataset generation for cybersecurity research. The objective is not to replicate any specific dataset, but to assess whether LLMs can closely approximate the statistical, structural, and semantic properties required for realistic security experimentation. In essence, our approach explores the potential of LLMs to enable testbed-free dataset generation, thereby reducing the cost, complexity, and operational overhead associated with traditional data collection infrastructures. To demonstrate and validate the proposed methodology, we employ the well-established AWID3~\cite{AWID3} intrusion detection benchmark as a representative use case. Namely, we employ a multi-level validation framework combining global similarity metrics, per-feature statistical analysis, dimensionality reduction, and cross-domain classification. Overall, this paper positions LLMs as a new class of controlled synthetic data generators. The results demonstrate that, when carefully constrained and validated, LLMs can generate structured datasets that support realistic experimentation while reducing reliance on costly and difficult real-world data collection.

The rest of the paper is structured as follows. The next section presents the proposed methodology for LLM-driven synthetic dataset generation, detailing the knowledge extraction process, the controlled multiphase generation pipeline, and the validation strategy. Section~\ref{S:res} reports the experimental results, including statistical similarity analysis, structural comparisons, and cross-domain classification performance across multiple LLMs and learning models. Section~\ref{S:insights} discusses the key strengths, limitations, and future work directions as derived from the empirical findings. Section~\ref{S:relatedwork} reviews related work on synthetic data generation and LLM applications in cybersecurity. The last section concludes the study and outlines directions for future research.

\section{Methodology}
\label{S:Methodology}

\begin{figure*} [!ht]
    \centering
    \includegraphics[width=1\linewidth]{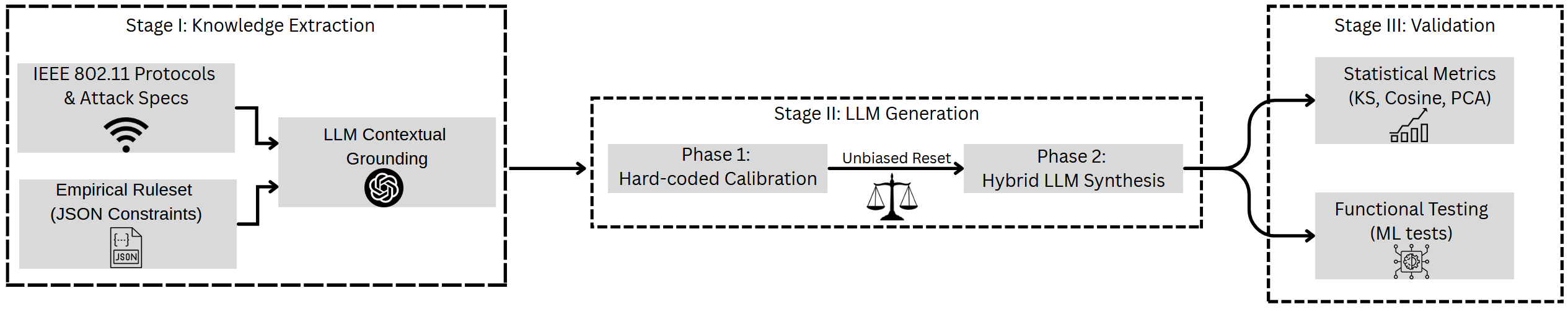}
    \caption{The proposed three-stage testbed-free generation pipeline.}
    \label{F:Methodology}
\end{figure*}

This section describes the methodology followed for generating synthetic datasets by combining natural language descriptions with explicit statistical constraints, using LLMs as controlled knowledge-to-data engines. Our objective is to assess whether LLM-generated data can accurately approximate the statistical distributions, structural dependencies, and semantic properties of well-established security datasets without direct access to the original data, obviating the need for implementing a testbed. The methodology presented in Figure~\ref{F:Methodology} follows a three-stage pipeline: (i) knowledge extraction, (ii) synthetic dataset generation, and (iii) validation. Note that, despite a specific intrusion detection dataset being used as a case study, the methodology itself is dataset-agnostic and can apply to any structured dataset.

As mentioned in Section~\ref{S:intro}, we select the AWID3 dataset~\cite{AWID3} as a representative case study. AWID3 provides a comprehensive collection of real-world IEEE~802.11 traffic traces, including both normal and malicious activity, captured in a controlled Wi-Fi testbed. Namely, the dataset targets IEEE~802.11 wireless traffic, which underpins the most widely deployed wireless communication protocol, not only in Small Office/Home Office (SOHO) and enterprise environments, but increasingly also in industrial ecosystems~\cite{kampourakis2025numeris}. It comprises four primary traffic classes, i.e., impersonation, injection, flooding, and normal traffic, including features extracted from Wi-Fi frames at both the physical (PHY) (e.g., signal strength, frame duration, transmission rate) and the Media Access Control (MAC) layers (e.g., frame type, source and destination MAC addresses, sequence numbers). Each instance is explicitly labeled, providing a categorical ground truth for supervised learning and systematic evaluation.

In summary, AWID3 exhibits rich semantic structure through numerous tightly constrained PHY and MAC layer features, making it a demanding test case for structured synthetic data generation. Our selection is further supported by the fact that the AWID family of datasets is widely adopted by the research community as a reference benchmark for Wi-Fi intrusion detection and, to the best of our knowledge, represents one of the most contemporary publicly available Wi-Fi security datasets, as it includes attack scenarios targeting modern IEEE~802.11 deployments and protection mechanisms aligned with the newest WPA3 certification, e.g., deauthentication attacks against the mandatory Protected Management Frames (PMF) scheme. Consequently, demonstrating the applicability of the proposed methodology in this semantically rich and protocol-constrained setting provides indirect evidence of its potential generalizability and extensibility to other security, but not limited to, datasets and case studies involving different protocols, domains, and feature semantics. 

\subsection{Knowledge Extraction}
\label{SS:knowledge:extr}

To initiate the synthetic generation process, we first construct natural language descriptions that summarize the operational characteristics and attack behaviors reflected in the AWID3 datasets, as seen at the leftmost side of figure~\ref{F:Methodology}. These descriptions are based on: i) the official IEEE 802.11 protocol documentation~\cite{80211}, ii) the nature and mechanism of each attack type, iii) information regarding the AWID3 dataset, involving class distributions, feature explanation, and utilized testbed description, iv) observed feature patterns associated with normal and anomalous activity, and v) statistical rulesets defining feature-wise probabilities, dependencies, and deterministic constraints empirically derived from the real dataset. These textual summaries are either directly extracted from existing documentation or manually composed, encapsulating the semantics of the original dataset. Additionally, numerical data are directly incorporated into the generation process. 

In detail, the LLM was provided with the following materials as attachments or manually crafted contextual grounding before generation. Specifically, the IEEE 802.11–2020 standard~\cite{80211}, to understand MAC and PHY layer structures, valid type–subtype mappings, and protocol constraints. Two academic papers~\cite{zargar2013survey,aminanto2016detecting} that describe flooding and impersonation attack types, including their observable network characteristics. The AWID3 dataset documentation~\cite{AWID3} and related publications~\cite{AWID3,chatzoglou2022pick} that explain the testbed setup, feature definitions, and class distributions. A manually formed ruleset, in JavaScript Object Notation (JSON) format, containing feature-wise distributions, probabilities, and constraints derived from empirical AWID3 analysis, as outlined in Appendix~\ref{A:ruleset:outline}. Regarding the ruleset, since the AWID3 dataset comprises IEEE 802.11 features characterized by high variability, we found that it is nearly impractical for the LLM to train solely on natural language descriptions. Therefore, we compiled complementary statistical and numeric data, necessary for the LLM generator to capture the dataset's structure and feature patterns. The data were derived from a 100k sample of the original AWID3 dataset by observing the range of allowed values for each feature, as well as the value dependencies that they may have.

\subsection{Synthetic Dataset Generation}

Following the preparation of the reference material and rulesets, as described in the knowledge extraction step~\ref{SS:knowledge:extr}, the LLM was interactively used as a controlled synthetic data generator. That is, the model was not fine-tuned or retrained; instead, it was conditioned through iterative prompt engineering. Subsequently, the dataset generation was performed in two distinct phases, as observed in the middle part of figure~\ref{F:Methodology}.

\noindent \textbf{\textit{Phase 1} — Hard-coded statistical ruleset:} In this phase, the model was provided with a preliminary ruleset in which each AWID3 feature had explicitly defined discrete values and associated occurrence probabilities. The LLM parsed these distributions directly from the JSON file and generated one million fully rule-compliant synthetic samples, ensuring that label counts, probabilistic restrictions (quotas), and protocol invariants were strictly satisfied. Note that this phase served as a reverse-engineering action, where the AWID3 dataset's statistical behavior was deconstructed into deterministic numerical patterns. Importantly, we calibrate the ruleset to achieve stability and fidelity across all major metrics, namely, precision, recall, F1-score, and accuracy, targeting a performance baseline where classifiers trained on \textit{Phase 1} data achieve at least 75\% of the F1-score observed when trained on the original AWID3 training set. We deliberately omit reporting the results of this phase, as they do not reflect autonomous LLM-based data generation, but rather the model's ability to strictly follow predefined instructions and constraints. Therefore, through \textit{Phase 1} and the hard-coded ruleset definition and calibration, we establish an unambiguous baseline from which a more descriptive representation is later derived in \textit{Phase 2}.

\noindent \textbf{\textit{Phase 2} — Descriptive and statistical hybrid ruleset:} Building upon \textit{Phase 1}, the LLM was supplied with a more descriptive version of the ruleset, containing fewer explicit probabilities and richer natural-language specifications of feature behavior. Notably, the same knowledge-base attachments mentioned in Section~\ref{SS:knowledge:extr} were reused. This phase tested the LLM's ability to infer missing quantitative relationships from descriptive context while maintaining IEEE 802.11–compliant feature logic and realistic statistical distributions. Importantly, to eliminate any potential memory persistence or model bias between phases, all intermediate data, prompts, and cached model states from \textit{Phase 1} were fully cleared. In other words, \textit{Phase 2} was initiated using a fresh session and a separate model instance, ensuring complete contextual isolation and unbiased generative behavior. Essentially, \textit{Phase 2} represented a controlled relaxation of the deterministic constraints defined in \textit{Phase 1}, demonstrating the LLM's capacity for generalized reasoning over structured, partially specified statistical systems. 

During both phases, the LLM was explicitly instructed to generate tabular data conforming exactly to the refined 16-feature AWID3 schema proposed in~\cite{chatzoglou2022pick}, follow deterministic sampling logic derived from the ruleset, outlined in Appendix~\ref{A:ruleset:outline}, adhering to a hierarchical sampling logic that prioritized protocol-specific locks, followed by management and overall traffic quotas, and finally post-flag categorical distributions, produce outputs in Comma-Separated Values (CSV) format, ensuring numeric fields remain within IEEE 802.11–valid ranges, and validate the resulting feature distributions and enforce the same class (highly imbalanced) ratios as in AWID3, i.e., Normal = 97\%, Flooding = 2\%, Impersonation = 1\%. Each dataset was generated directly within the LLM environment under strict rule enforcement. Namely, intermediate validation logs were printed to verify compliance with quotas, distributions, and feature dependencies. The final one-million–sample dataset from each phase was exported in CSV format for downstream training and evaluation. Note that all analyses, comparisons, and results presented in this study refer exclusively to datasets generated during \textit{Phase 2}. Recall that \textit{Phase 1} served solely as a controlled calibration stage to construct the statistical foundation required for \textit{Phase 2} synthesis. It is also important to note that we confirmed that the LLM had no direct access to the real AWID3 dataset, as seen below.

\begin{mdframed}[
  backgroundcolor=gray!10,
  linecolor=black,
  linewidth=0.4pt,
  nobreak=true
]
\label{Fr:Statement}
\noindent\textbf{Statement on prior access and data independence (ChatGPT).}
Before this study, I did not have access to the AWID3 dataset, nor to any raw samples, packet traces, or feature-level numerical data derived from it. I do not retain or retrieve specific benchmark datasets as part of my general training. Any AWID3-specific understanding used during generation was acquired solely from documentation, rulesets, and descriptive materials explicitly provided within this interaction. No original data samples were accessed, recalled, or reconstructed, and all generated outputs are data-independent while remaining statistically aligned by design.
\end{mdframed}

\subsection{Validation}
\label{SS:validation}

This section details the evaluation process followed to measure the fidelity and practical validity of the LLM-generated datasets. To this end, we employ five complementary evaluation strategies: (i) cosine similarity and (ii) Euclidean distance, (iii) the Kolmogorov–Smirnov, (iv) principal component analysis, and (v) cross-domain classification using the LightGBM, Random Forest, and MLP classifiers to validate the semantic consistency of the synthetic data vis-\`a-vis the real AWID3 dataset. Each of the validation methods we utilized is briefly described to facilitate the readers' comprehension and subtle understanding of the following sections.

\noindent \textbf{Cosine Similarity.} Cosine similarity measures the directional alignment between two datasets represented as feature vectors. In this context, $\mathbf{A}$ corresponds to the feature space of the real AWID3 dataset, and $\mathbf{B}$ to the feature space of the synthetic dataset, as seen in equation~\ref{eq:cosinesim}. A cosine similarity value close to $1$ indicates that the synthetic dataset exhibits nearly identical directional structure in feature space to the real AWID3 data, implying strong global correspondence in multivariate patterns.

\begin{equation}
\label{eq:cosinesim}
    S_{C}(A,B) = 
    \cos(\theta) = 
    \frac{\mathbf{A} \cdot \mathbf{B}}
    {\|\mathbf{A}\| \, \|\mathbf{B}\|} = 
    \frac{\sum\limits_{i=1}^{n} A_i B_i}
    {\sqrt{\sum\limits_{i=1}^{n} A_i^2} \,
     \sqrt{\sum\limits_{i=1}^{n} B_i^2}}
\end{equation}

\noindent \textbf{Euclidean Distance.} Euclidean distance quantifies the geometric separation between the two datasets in feature space. Here, lower distances correspond to smaller deviations in the numerical representation of the underlying feature distributions, as detailed in equation~\ref{eq:euclideandis}. In the AWID3 context, a smaller $D_E$ indicates that the synthetic samples preserve the same feature magnitudes and scaling relationships as the original network traffic, reflecting accurate replication of the global numeric landscape.

\begin{equation}
\label{eq:euclideandis}
    D_{E}(A,B) = 
    \sqrt{\sum\limits_{i=1}^{n} (A_i - B_i)^2}
\end{equation}

\noindent \textbf{Kolmogorov–Smirnov (KS) statistic.} The KS statistic measures the maximum difference between the empirical cumulative distribution functions (ECDFs) of corresponding features from the real and synthetic datasets. In equation~\ref{eq:ksstat}, $F_{1}(x)$ and $F_{2}(x)$ denote the ECDFs of a given feature (e.g., signal strength, frame duration, or channel frequency) in the real and synthetic datasets, respectively. A lower $D_{\text{KS}}$ implies that the per-feature distribution of the synthetic data closely follows that of the real AWID3 dataset, confirming local statistical fidelity.

\begin{equation}
\label{eq:ksstat}
    D_{\text{KS}} = 
    \sup_{x} \left| F_{1}(x) - F_{2}(x) \right|
\end{equation}

\noindent \textbf{Principal Component Analysis (PCA).} PCA is used to project both datasets into a shared low-dimensional subspace for structural comparison, as seen in~\ref{eq:pca}. When applied to the AWID3 and synthetic datasets, PCA visualizations reveal whether the two occupy similar regions in variance space. Overlapping cluster structures in the projection indicate that the LLM successfully preserved the global covariance structure and inter-feature dependencies underlying the real network traffic.

\begin{equation}
\label{eq:pca}
    Z = XW, \quad 
    W = \arg\max_{W^{T}W = I} \, \mathrm{Var}(XW)
\end{equation}

\noindent \textbf{Cross-domain classification evaluation.} To evaluate the semantic and functional validity of the generated datasets, the supervised Random Forest and LightGBM classifiers and a Multi-Layer Perceptron (MLP) feedforward network were trained exclusively on the synthetic data and tested on the real AWID3 samples. This cross-domain setup assesses whether the synthetic data preserves the same discriminative structure and decision boundaries that separate normal, flooding, and impersonation traffic in the real dataset.

Model performance is measured using standard classification metrics, including precision, recall, F1-score, accuracy, and the Confusion Matrix (CM) of Appendix~\ref{A:conf:matrix}. These metrics jointly capture the classifiers' ability to identify both normal and attack traffic patterns correctly. High F1 and accuracy values, accompanied by balanced precision–recall scores across classes, indicate that the synthetic dataset effectively replicates the behavioral characteristics and feature relationships of the original AWID3 data~\cite{kampourakis2025balancing}. The confusion matrix further provides a class-level breakdown of predictions, confirming whether the model trained on synthetic data generalizes to real network conditions without significant degradation in detection capability.

\section{Results}
\label{S:res}

This section presents the results of the validation experiments, described in Section~\ref{SS:validation}. Recall that the objective is to quantify how effectively LLMs can replicate the statistical, structural, and behavioral characteristics of the AWID3 dataset while maintaining feature interdependencies and attack-class discriminability. For our validation benchmark, we utilized the interface of four state-of-the-art LLMs: ChatGPT-5, Gemini 2.5 Pro, Claude Opus 4.1, and Qwen3-Max. Data collection and generative experiments were concluded by November 2025. Each LLM was tasked with generating synthetic AWID3-like datasets based on the identical contextual inputs, detailed in Section~\ref{SS:knowledge:extr}. Likewise, all models operated under the same schema and ruleset constraints to ensure fair comparability of the generated data.

\begin{figure*}[!ht]
\centering
\captionsetup[subfigure]{justification=centering}
\subfloat[\textbf{ChatGPT--5}]{
    \includegraphics[width=0.42\textwidth]{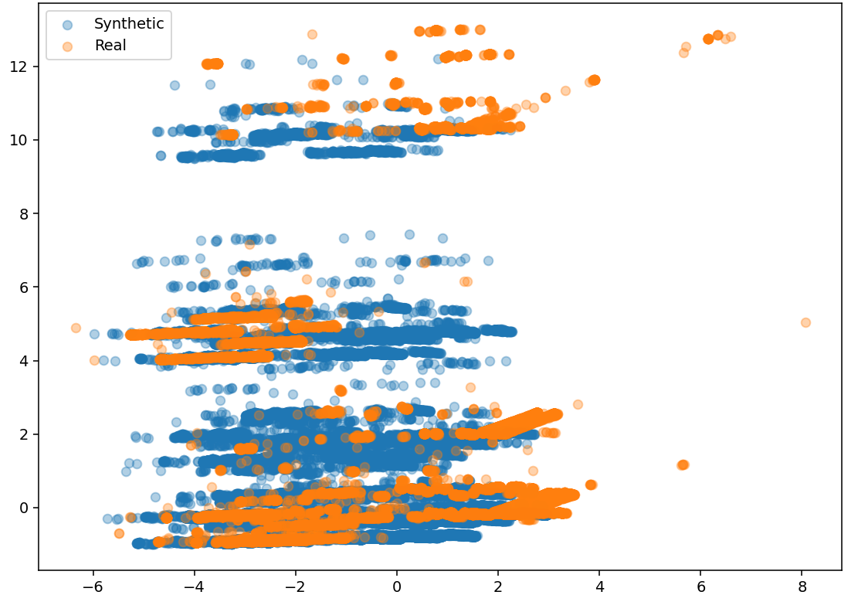}
}
\hfill
\subfloat[\textbf{Gemini 2.5 Pro}]{
    \includegraphics[width=0.42\textwidth]{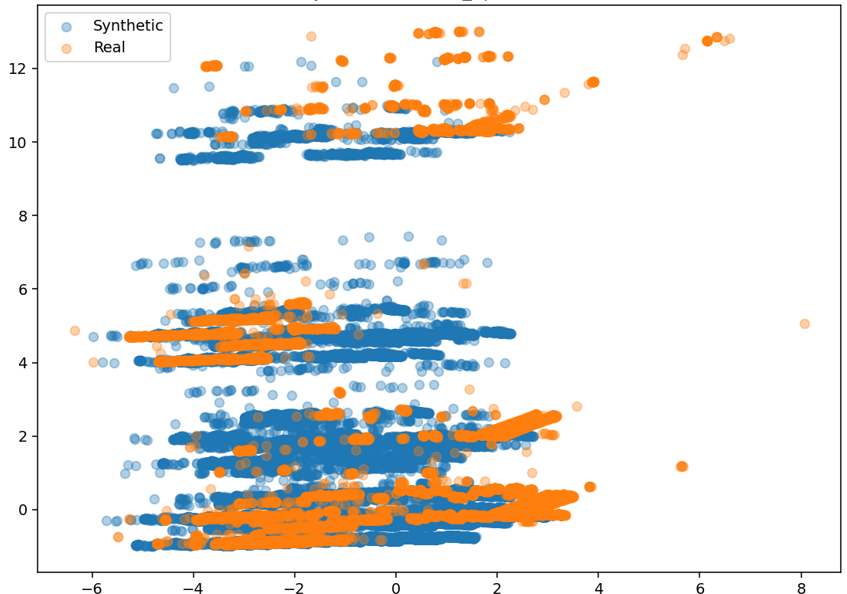}
}
\\[2ex]
\subfloat[\textbf{Claude Opus 4.1}]{
    \includegraphics[width=0.42\textwidth]{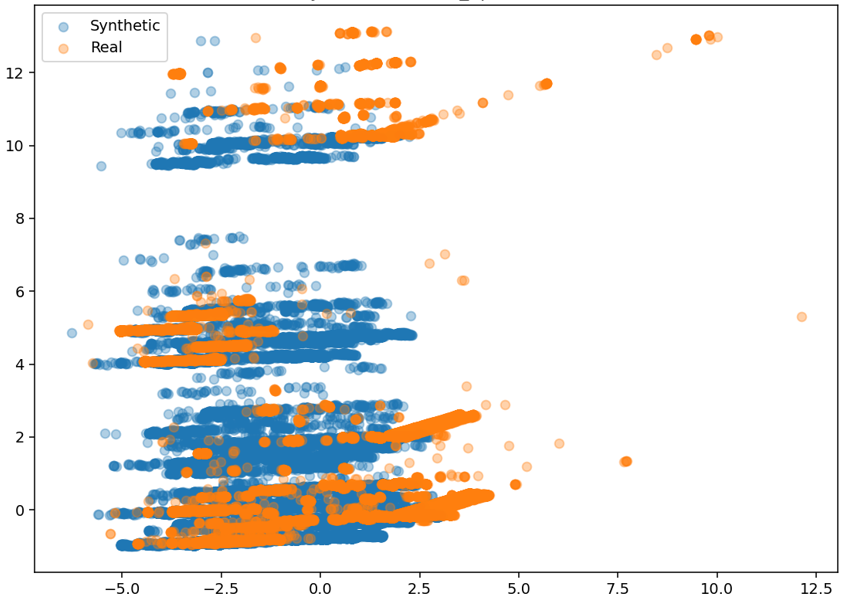}
}
\hfill
\subfloat[\textbf{Qwen3--Max}]{
    \includegraphics[width=0.42\textwidth]{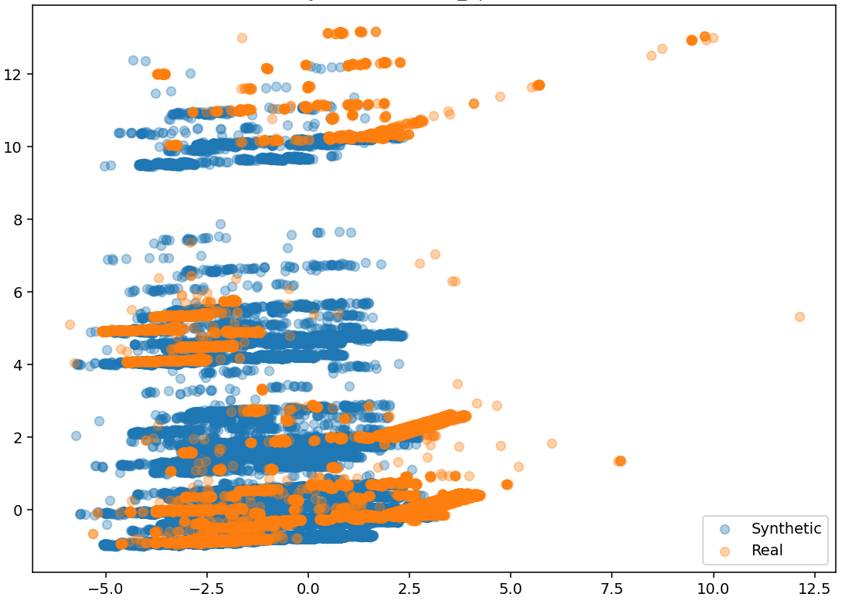}
}
\caption{Two-dimensional PCA projections comparing real (orange) and LLM-generated (blue) AWID3 data across all four models.}
\label{F:PCA}
\end{figure*}

The generated datasets were quantitatively compared against the real AWID3 dataset using the defined validation framework. Cosine similarity and Euclidean distance were employed to assess global structural alignment between the synthetic and real datasets, while the KS statistic was computed feature-wise to evaluate the fidelity of each feature’s statistical distribution. Dimensional and structural consistency were further examined through PCA scatter plots, visualizing the overlap between real and synthetic samples in reduced feature space, as depicted in~\ref{F:PCA}. Finally, the Random Forest and LightGBM classifiers, as well as the MLP neural network, were trained exclusively on synthetic data and evaluated on real AWID3 samples, with performance measured through accuracy, precision, recall, F1-score, and confusion matrix analysis, as previously delineated in Section~\ref{SS:validation}.

\begin{table*}[htpb]
\centering
\caption{Global similarity between real and synthetic datasets. Each entry reports Mean / Median / Std. Best values are highlighted in blue, worst in red.}
\label{T:Cos_Euc}
\begin{tabular}{l|ccc|ccc}
\hline
\textbf{Model} & \multicolumn{3}{c|}{\textbf{Euclidean Distance}} & \multicolumn{3}{c}{\textbf{Cosine Similarity}} \\
 & \textbf{Mean} & \textbf{Median} & \textbf{Std} & \textbf{Mean} & \textbf{Median} & \textbf{Std} \\ \hline
ChatGPT\textendash 5     & \textcolor{blue}{1005.85} & \textcolor{blue}{358.87} & \textcolor{blue}{1094.50} & \textcolor{blue}{0.97937} & 0.99851 & \textcolor{blue}{0.04438} \\
Gemini 2.5 Pro           & 1019.60 & 372.14 & \textcolor{red}{1125.30} & 0.97783 & \textcolor{red}{0.99762} & 0.04522 \\
Claude Opus 4.1          & \textcolor{red}{1059.70} & \textcolor{red}{427.58} & 1100.17 & \textcolor{red}{0.97678} & 0.99857 & \textcolor{red}{0.04536} \\
Qwen3\textendash Max     & 1051.06 & 386.03 & 1100.46 & 0.97713 & \textcolor{blue}{0.99862} & 0.04534 \\
\hline
\end{tabular}
\end{table*}

Table~\ref{T:Cos_Euc} summarizes the global similarity between the real and LLM-generated AWID3 datasets, evaluated using Euclidean distance and cosine similarity. As observed from the table, ChatGPT-5 achieved the closest alignment with the real dataset, exhibiting the lowest Euclidean mean (1005.85), median (358.87), and standard deviation (1094.50), as well as the highest cosine mean (0.97937) and lowest cosine variability (0.04438). Qwen3-Max followed closely, yielding the highest cosine median (0.99862), indicating strong typical directional similarity, though with slightly higher Euclidean distance than ChatGPT-5. Gemini 2.5 Pro produced competitive averages but showed the largest Euclidean spread, suggesting a few outlier deviations. Claude Opus 4.1 had the weakest overall performance, with the highest Euclidean mean and lowest cosine mean, reflecting lower fidelity to the real data distribution. In general terms, the results showcase that all four LLMs successfully captured the general structural and statistical patterns of the AWID3 dataset, with ChatGPT-5 demonstrating the most accurate and stable reproduction of both numerical and geometric relationships between features.

\begin{table*}[!ht]
\centering
\caption{Top-3 per-feature KS distances (lower is better). Each cell shows \emph{feature} (KS).}
\label{T:KS}
\resizebox{\textwidth}{!}{%
\begin{tabular}{l|lll}
\hline
\textbf{Model} &
\multicolumn{1}{c}{\textbf{1st}} &
\multicolumn{1}{c}{\textbf{2nd}} &
\multicolumn{1}{c}{\textbf{3rd}} \\ \hline
ChatGPT--5      & \textit{wlan.duration} (0.29439) & \textit{frame.len} (0.24179) & \textit{radiotap.dbm\_antsignal} (0.10986) \\
Gemini 2.5 Pro  & \textit{wlan.duration} (0.31402) & \textit{frame.len} (0.26458) & \textit{radiotap.dbm\_antsignal} (0.11692) \\
Claude Opus 4.1 & \textit{frame.len} (0.14352)     & \textit{wlan.duration} (0.12041) & \textit{radiotap.dbm\_antsignal} (0.10947) \\
Qwen3--Max      & \textit{frame.len} (0.15071)     & \textit{wlan.duration} (0.12054) & \textit{radiotap.dbm\_antsignal} (0.10971) \\
\hline
\end{tabular}%
}
\end{table*}

Table~\ref{T:KS} reports the three features with the highest KS distances for each model, indicating where the largest distributional deviations occur between the synthetic and real AWID3 data. The features \textit{wlan.duration}, \textit{frame.len}, and \textit{radiotap.dbm\_antsignal} consistently appear among the top KS values across all models, suggesting that these attributes exhibit the greatest variability and are therefore the most challenging to replicate accurately. However, this behavior is somewhat expected. Namely, the \textit{wlan.duration} and \textit{frame.len} features are inherently bursty, influenced by traffic type and transmission conditions, while \textit{radiotap.dbm\_antsignal} reflects real-world signal fluctuations due to channel interference and antenna dynamics. Capturing these stochastic PHY layer properties is difficult for a pure LLM-driven generator lacking real-time temporal context. Interestingly, while ChatGPT-5 leads in global metrics, Claude Opus 4.1 and Qwen3-Max demonstrate superior local fidelity for these specific protocol-heavy features, achieving lower KS distances for \textit{frame.len} and \textit{wlan.duration}. This suggests that despite the difficulty in modeling signal variance, certain models may be more adept at capturing rigid, discrete protocol constraints.

Figure~\ref{F:PCA} presents the two-dimensional PCA scatter plots for real versus LLM-generated AWID3 data. Across all models, the synthetic samples exhibit strong spatial overlap with the real dataset, indicating that the generated features preserve the overall variance structure and inter-feature relationships. The tight clustering of attack and normal classes further demonstrates that all four LLMs captured both the statistical and geometric structure of the underlying IEEE 802.11 traffic patterns. Note that the minor dispersion observed in the synthetic points reflects variability introduced during generative sampling rather than systematic bias. For the cross-domain classification process, we selected LightGBM, Random Forest, and MLP classifiers, covering gradient-boosting, ensemble-based, and neural architectures. Specifically, LightGBM provides a high-capacity gradient-boosting baseline sensitive to feature interactions, Random Forest offers a robust ensemble learner resilient to distributional noise, and MLP evaluates whether the synthetic data supports representation learning under neural architectures. 

\begin{table*}[htpb]
\centering
\caption{Cross-domain classification results (train on synthetic, test on real). Best values are highlighted in blue, worst in red.}
\label{T:cross_domain_metrics}
\begin{tabular}{l|rrrr}
\hline
\rowcolor{lightgray}
\multicolumn{5}{c}{\textbf{LightGBM}} \\
\hline
\textbf{Model} & \textbf{Precision} & \textbf{Recall} & \textbf{F1-Score} & \textbf{Accuracy} \\
\hline
ChatGPT\textendash 5     & 0.9399 & 0.9620 & 0.9487 & 0.9945 \\
Gemini 2.5 Pro           & \textcolor{red}{0.9372} & \textcolor{red}{0.9593} & \textcolor{red}{0.9460} & \textcolor{red}{0.9942} \\
Claude Opus 4.1          & 0.9361 & \textcolor{blue}{0.9780} & 0.9543 & 0.9948 \\
Qwen3\textendash Max     & \textcolor{blue}{0.9404} & 0.9754 & \textcolor{blue}{0.9558} & \textcolor{blue}{0.9950} \\
\hline
\rowcolor{lightgray}
\multicolumn{5}{c}{\textbf{Random Forest}} \\
\hline
\textbf{Model} & \textbf{Precision} & \textbf{Recall} & \textbf{F1-Score} & \textbf{Accuracy} \\
\hline
ChatGPT\textendash 5     & 0.9423 & \textcolor{red}{0.8961} & \textcolor{red}{0.9090} & \textcolor{red}{0.9927} \\
Gemini 2.5 Pro           & 0.9425 & 0.8993 & 0.9113 & 0.9928 \\
Claude Opus 4.1          & \textcolor{red}{0.9416} & 0.9006 & 0.9116 & 0.9928 \\
Qwen3\textendash Max     & \textcolor{blue}{0.9469} & \textcolor{blue}{0.9256} & \textcolor{blue}{0.9299} & \textcolor{blue}{0.9941} \\
\hline
\rowcolor{lightgray}
\multicolumn{5}{c}{\textbf{MLP}} \\
\hline
\textbf{Model} & \textbf{Precision} & \textbf{Recall} & \textbf{F1-Score} & \textbf{Accuracy} \\
\hline
ChatGPT\textendash 5     & \textcolor{blue}{0.8410} & 0.7772 & \textcolor{blue}{0.8022} & \textcolor{blue}{0.9842} \\
Gemini 2.5 Pro           & 0.8410 & 0.7768 & 0.8020 & 0.9841 \\
Claude Opus 4.1          & 0.7535 & \textcolor{red}{0.7447} & \textcolor{red}{0.6733} & \textcolor{red}{0.9691} \\
Qwen3\textendash Max     & \textcolor{red}{0.7180} & \textcolor{blue}{0.9503} & 0.7880 & 0.9759 \\
\hline
\end{tabular}
\end{table*}

Table~\ref{T:cross_domain_metrics} presents the cross-domain classification performance obtained by training each classifier exclusively on synthetic data and evaluating it on real samples. While accuracy is reported for the sake of completeness, it is a known poor indicator of performance in highly imbalanced environments such as AWID3, where the majority class (Normal) accounts for 97\% of the instances. Consequently, our analysis primarily focuses on the F1-score and Recall, which more accurately reflect the models' ability to identify the minority attack classes (Flooding and Impersonation). Overall, LightGBM achieves the strongest and most consistent results across all LLM-generated datasets, with Qwen3–Max yielding the highest F1-score (0.9558) and accuracy (0.9950), while all LightGBM configurations maintain precision and recall above 0.93 and 0.95, respectively, demonstrating strong preservation of discriminative structure. Random Forest exhibits slightly lower recall and F1-scores compared to LightGBM, but maintains stable accuracy levels close to 0.993 across all models, reflecting robustness to moderate distributional variability in the synthetic data. In contrast, MLP shows noticeably lower and more variable performance, with the lowest F1-score of 0.6733 and accuracy dropping to 0.9691 in the weakest case, indicating higher sensitivity to residual statistical mismatches and highlighting the greater difficulty of transferring neural representations from synthetic to real domains. Overall, the results indicate that all evaluated LLMs were able to generate synthetic datasets that closely preserve the statistical, structural, and behavioral characteristics of the AWID3 dataset. Each model effectively captured the underlying traffic dynamics and attack behavior patterns associated with flooding and impersonation scenarios. Minor deviations were observed in high-variance features such as signal strength and frame duration, yet the overall alignment across distributions and classification performance confirms that the generated data maintains strong fidelity to the real AWID3 dataset.

\section{Strengths, Limitations, \& Future Work}
\label{S:insights}

This section provides several key insights regarding the strengths, limitations, and future direction of using LLMs for synthetic dataset generation in cybersecurity.

\subsection{Strengths}

The results of Section~\ref{S:res} demonstrate that LLMs, when grounded in protocol-specific knowledge and explicit statistical constraints, can generate synthetic network traffic datasets that preserve key statistical distributions, structural dependencies, and semantic characteristics of real data. This holds even for the demanding IEEE~802.11 use case considered in this study. Specifically, across global similarity metrics, feature-wise statistical tests, PCA-based structural comparisons, and cross-domain classification, the synthetic datasets demonstrated strong alignment with the AWID3 benchmark, as summarized in Tables~\ref{T:Cos_Euc} and~\ref{T:cross_domain_metrics} and Figure~\ref{F:PCA}. Notably, the efficacy of gradient-boosting and ensemble learners in the cross-domain task suggests that LLMs successfully encode the underlying decision logic of the network protocol. Importantly, models trained exclusively on synthetic data generalized effectively to real traffic, particularly for gradient-boosting and ensemble-based learners (LightGBM and Random Forest), indicating that LLM-generated datasets can support realistic IDS experimentation when access to real-world traces is limited or impractical.

Beyond performance, a key strength of the proposed methodology lies in its dataset-agnostic and controllable design. While AWID3 serves as a representative case study, the methodology itself is independent of any specific dataset and can be applied to other protocols, domains, and feature schemas. Unlike many existing LLM-based synthetic data approaches, this work does not rely on fine-tuning or direct exposure to raw samples, thereby reducing the risk of data leakage and making the approach suitable for compliance-sensitive environments. Moreover, the use of explicit rulesets and protocol constraints provides a high degree of explainability, enabling targeted dataset modification, principled debugging, and semantic extension of attack behaviors without requiring new data collection campaigns.

\subsection{Limitations}

Despite the strong overall alignment, several limitations remain. First, the current generation process focuses on tabular fidelity and does not explicitly model temporal continuity or long-range dependencies, which may limit realism for sequence-based or flow-level IDS tasks. In other words, this limitation is evident for high-variance features influenced by environmental and temporal dynamics, such as signal strength and bursty duration or length patterns, as indicated by the top KS distances in table~\ref{T:KS}. Second, although the approach eliminates the need for testbeds, it still requires non-trivial manual effort to construct statistical rulesets and validate constraints, which may introduce bias or limit scalability without further automation.

Finally, cross-domain generalization performance varies across learning models, with neural architectures exhibiting higher sensitivity to residual distributional mismatches than tree-based methods, indicating that synthetic-to-real transferability is not uniformly guaranteed. The performance gap observed between tree-based learners and the MLP architecture suggests a fundamental difference in how these models consume synthetic data. While LLMs effectively replicate the discrete logical thresholds, e.g., protocol-specific rules, used by LightGBM to partition data, they struggle to synthesize the smooth statistical manifolds and stochastic noise that neural networks rely on for feature representation. This lack of stochastic nuance in LLM-generated data highlights a key challenge in achieving high-fidelity transferability for deep learning architectures.

\subsection{Future Work}

Building upon the empirical findings and identifying the remaining gaps, several directions emerge for future research. One promising avenue involves the integration of explicit temporal modeling to capture sequential dependencies and burst-level dynamics in network traffic, potentially through hybrid architectures that combine LLM-based constraint reasoning with Generative Adversarial Networks (GANs) or specialized time-series simulators. Moreover, the automation of the extraction and refinement of statistical rulesets from limited real data or expert feedback could further improve scalability and reduce manual effort. Additionally, extending the evaluation to other protocols, network environments, and attack classes would help assess the generalizability of the approach. Finally, future studies could explore adaptive generation strategies that iteratively refine synthetic datasets based on downstream detection performance, enabling closed-loop optimization of synthetic data quality for specific security tasks.

\subsubsection*{Work in Progress.} A significant advantage of the proposed LLM-driven methodology is its potential for synthesizing zero-day attack patterns, i.e., malicious behaviors that are not present in the original AWID3 training set. Traditional synthetic generators are often limited to the statistical bounds of the historical data they were trained on. In contrast, the reasoning capabilities of LLMs allow them to extrapolate from abstract security concepts and protocol documentation to generate novel attack signatures. In the context of Wi-Fi-specific attacks, such novel and semantically complex attacks include new variants of the well-known KRACK attack~\cite{vanhoef-crack-new}, fragmentation attacks~\cite{vanhoef-frag}, and side-channel attacks~\cite{vanhoef-dragonblood}. To evaluate this, we define a \textit{semantic injection} process where the LLM is provided with a natural language description of a hypothetical or emerging vulnerability, e.g., a novel WPA3 handshake exhaustion. By leveraging its internal knowledge of the IEEE 802.11 protocol stack, the LLM projects these semantic descriptions into the 16-feature space used in our schema. This enables a synthetic-to-zero-day validation scenario. In this setup, we assess whether a classifier trained on LLM-synthesized data can detect variants of attacks that were deliberately excluded from the original training baseline. This capability suggests that LLMs can serve not only as data replicators but as proactive security engines, allowing researchers to generate labeled datasets for emerging threats before they are even observed in a physical testbed or captured in the wild.

\noindent \textbf{\textit{Sneak Peek into Preliminary Results.}} As an initial validation of the proposed semantic injection framework for zero-day attack detection, we conducted a preliminary cross-domain experiment in which a LightGBM classifier was trained on a synthetically generated dataset containing semantically defined attack patterns derived from recently disclosed Block ACK (BA) and Block ACK Request (BAR) deauthentication attacks described in~\cite{Bl0ck} and evaluated on a disjoint real-data block. Specifically, both training and testing sets comprised 100k samples, with a class distribution of 70\% benign and 30\% attack traffic. Despite the absence of any real zero-day samples during training, the classifier achieved a precision of 0.7293, a recall of 0.7100, an F1-score of 0.7196, and an overall accuracy of 0.8339. Obviously, these results are notably lower than those observed for in-distribution attack scenarios, nonetheless, they indicate that LLM-synthesized semantic attack representations can induce transferable decision margins that generalize to previously unseen real-world behaviors. These early findings provide encouraging evidence that LLM-driven synthetic generation may support proactive IDS evaluation under zero-day conditions, motivating more extensive and systematic investigation.

\section{Related Work}
\label{S:relatedwork}

This section reviews prior work on synthetic data generation using LLMs, focusing on cybersecurity applications. 

In~\cite{Almorjan_2025}, the authors proposed an LLM-based scheme for generating synthetic textual cybersecurity datasets focused on Indicators of Compromise (IoCs) in social media content. Specifically, they fine-tuned a GPT-3.5 instance on real social media datasets and curated IoC knowledge, enabling the generation of labeled comments to emulate platform-specific writing styles. Subsequently, the synthetic data were evaluated by training conventional machine- and deep-learning classifiers to perform IoC classification on the generated text. Altogether, this study targeted unstructured text and Cyber Threat Intelligence (CTI)-style data, relying on LLM fine-tuning with direct access to real samples.

The work in~\cite{Goyal2025} presented a general-purpose LLM-based synthetic data generation platform, combining fine-tuned LLMs with differential privacy mechanisms to generate synthetic datasets across multiple domains, including cybersecurity. The proposed framework enabled users to either augment existing datasets or generate new tabular data from high-level descriptions, with privacy guarantees ensured through differential privacy techniques. The authors evaluated their scheme on small, generic tabular datasets, focusing on statistical similarity, privacy–utility trade-offs, and downstream machine-learning performance. Overall, the authors targeted domain-agnostic tabular data and relied on LLM fine-tuning and differential privacy.

The authors in~\cite{Galadima2024} explored the use of LLMs to generate synthetic cyber incident response process logs. Their approach augmented an existing incident response dataset by using few-shot prompting with ChatGPT and Gemini, guided by incident response playbooks and domain-expert refinement, to generate interconnected event logs and textual communication data that reflect incident response workflows. Overall, this study focused on process and communication logs with strong textual components and performed dataset augmentation using in-context examples.

In~\cite{patel2024datadreamer}, the authors introduced \textit{DataDreamer}, an open-source Python framework designed to support reproducible LLM-in-the-loop workflows, including synthetic data generation, dataset augmentation, and model fine-tuning, through prompt-based LLM generation and chained processing steps. Specifically, they focused on tooling and infrastructure, providing standardized abstractions for prompting, chaining multi-stage workflows, caching, reproducibility fingerprints, and publishing synthetic datasets and models, primarily targeting natural language processing and general-purpose machine-learning research.

Altogether, the body of work on LLM-driven synthetic dataset generation remains limited, particularly within the cybersecurity domain. Notably, all the previously discussed works emerged very recently (2024–2025), underscoring that this research direction is still in its early stages. Current efforts primarily focus on unstructured textual artifacts, e.g., CTI narratives and incident response logs, domain-agnostic tabular data, or example-driven dataset augmentation, often relying on fine-tuning or direct exposure to real datasets. In contrast to these existing efforts, our study addresses a largely unexplored paradigm: the generation of structured, protocol-constrained network traffic without the prerequisite of fine-tuning or direct access to raw samples. While current literature primarily addresses unstructured textual artifacts, e.g., CTI narratives, or general-purpose tabular data through example-driven augmentation, this work investigates the LLM's capacity to serve as an autonomous protocol-reasoning engine. By emphasizing statistical fidelity, protocol semantics, and cross-domain IDS generalization, this research shifts the focus from mere data replication to knowledge-driven synthesis, contributing one of the first systematic evaluations of LLMs as controlled generators for highly specific cybersecurity environments.

\section{Conclusions}
\label{S:conclusions}

This paper examined the feasibility of using LLMs as controlled generators of structured synthetic network traffic data for IDS research. Motivated by the high cost and logistical complexity of physical testbeds, we proposed a methodology that grounds LLMs in protocol semantics and statistical constraints, bypassing the need for fine-tuning or direct access to raw datasets. Our evaluation using the AWID3 benchmark across four state-of-the-art LLMs demonstrates that this knowledge-driven synthesis preserves the statistical and structural properties of real-world traffic. Notably, synthetic datasets achieved cosine similarity means up to 0.979 and enabled high-fidelity cross-domain detection. Gradient-boosting learners trained exclusively on synthetic data achieved F1-scores up to 0.956, demonstrating that LLMs can successfully encode the discriminative logic of network attacks. While minor discrepancies remain in capturing stochastic signal-layer noise, the results provide compelling evidence that LLMs can serve as a practical, testbed-free alternative for generating high-quality cybersecurity data. Ultimately, this work paves the way for a more agile research paradigm where realistic datasets are synthesized on-demand. This approach drastically accelerates the R\&D cycle for robust IDS by bypassing the resource-intensive bottlenecks of physical data collection, manual labeling, and the restrictive privacy regulations associated with sharing real-world network traces.

\appendix

\section{Confusion Matrix}
\label{A:conf:matrix}

\begin{figure*}[!ht]
\centering
\captionsetup[subfigure]{justification=centering}
\subfloat[\textbf{ChatGPT--5}]{
    \includegraphics[width=0.35\textwidth]{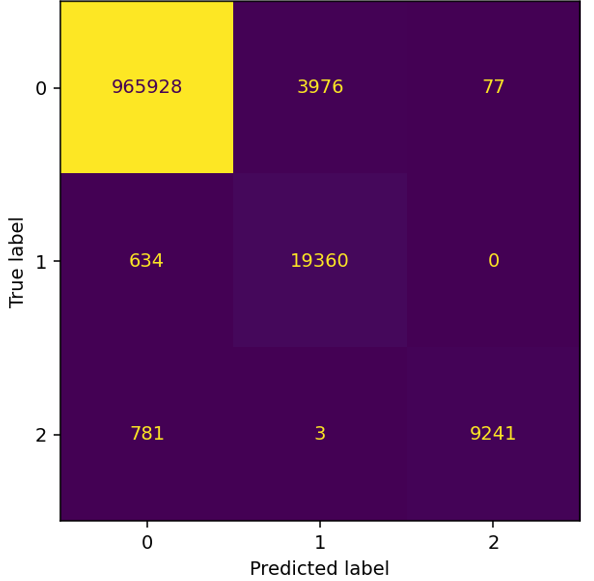}
}
\hfill
\subfloat[\textbf{Gemini 2.5 Pro}]{
    \includegraphics[width=0.35\textwidth]{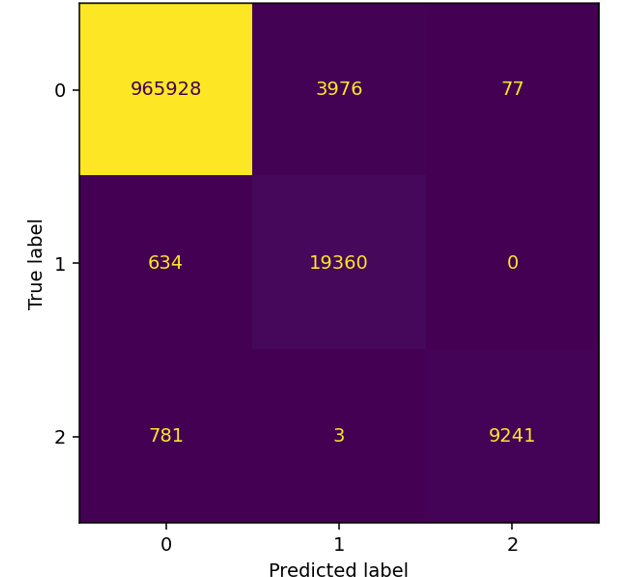}
}
\\[2ex]
\subfloat[\textbf{Claude Opus 4.1}]{
    \includegraphics[width=0.35\textwidth]{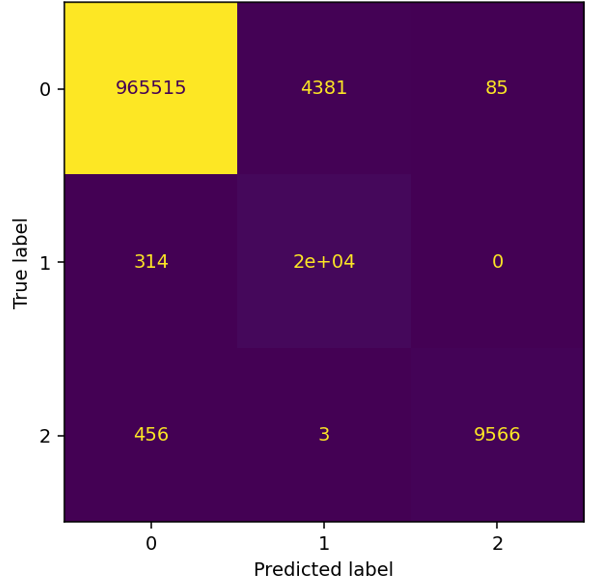}
}
\hfill
\subfloat[\textbf{Qwen3--Max}]{
    \includegraphics[width=0.35\textwidth]{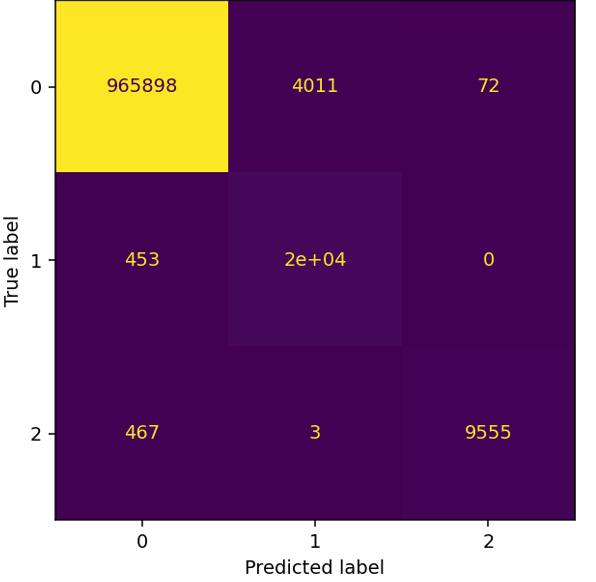}
}
\caption{Confusion matrices for the best-performing classifier (LightGBM) across all four LLM-generated datasets. All models demonstrate high true positive rates for both normal and attack classes.}
\label{F:CMs}
\end{figure*}

Figure~\ref{F:CMs} presents the CMs obtained from the cross-domain classification experiment, where a LightGBM model was trained on synthetic datasets and evaluated on real AWID3 samples. All four LLM-generated datasets demonstrate strong alignment with the real data, achieving high recognition accuracy across normal, flooding, and impersonation classes. Minor misclassifications occur primarily between normal and flooding traffic, which share overlapping temporal and signal characteristics. These results confirm that the generated datasets retain meaningful discriminative structure consistent with real network behaviors.

\section{Ruleset Outline}
\label{A:ruleset:outline}

The pseudocode in Listing~\ref{A:code} defines a rule-strict Wi-Fi traffic generator that produces exactly \textit{N} rows according to label priors and per-label distributions observed by the human expert in real network traces. For each label, it samples the core fields: type/subtype, channel frequency, and PHY flags, including \textit{radiotap.length}, RSSI, DS direction, \textit{frame.len} from discrete, label-specific supports, and \textit{wlan.duration}. Each row is validated against hard constraints, namely, subtype-to-type allowlists, band/flag consistency, DS rules, TSFT must be 0 when \textit{radiotap.length}=64, legal control durations, and supported \textit{frame.len}); otherwise it is rejected and resampled. After sampling, a quota stage applies: hard locks are enforced first, followed by \textit{wlan.fc.protected} and other flags (e.g., \textit{retry}, \textit{pwrmgt}, \textit{moredata}) that are matched to per-label targets, while never affecting guarded rows or forbidden control subtypes. Separating sampling, validation, and quota adjustment preserves target distributions and multi-field invariants. A single Random Number Generator (RNG) seed yields full reproducibility. Optional label-specific overrides (e.g., flooding: \textit{radiotap.length}=56 with \textit{radiotap.present.tsft}=1) can be applied before the global mixes to correct deviations without affecting other classes.

{\lstset{
  basicstyle=\ttfamily\tiny,
  frame=single,
  numbers=left,
  numbersep=6pt,
  xleftmargin=1.5ex,
  aboveskip=0.6em,
  belowskip=0.6em,
  breaklines=true,
  columns=fullflexible
}

{\lstset{basicstyle=\ttfamily\scriptsize,frame=single,breaklines=true,numbers=left,numbersep=6pt,xleftmargin=1.5ex}
\begin{lstlisting}[caption={Ruleset Pseudocode},label={A:code}]
# Globals: R ruleset; N rows; SEED; CLASSES={0,1,2}

PROCEDURE GENERATE(R,N,SEED):
  rng := InitRNG(SEED); C := R.label_distribution.counts
  assert sum(C.values)=N
  D := empty
  for label in CLASSES:
    while COUNT(D where Label=label) < C[label]:
      r := EMIT_ROW(R,label,rng); if CHECK(R,r): APPEND(D,r)
  D := ENFORCE_QUOTAS(R,D); assert EXACT_LABEL_COUNTS(D,C); return D

FUNCTION EMIT_ROW(R,label,rng) -> row:
  # Type/Subtype
  t  := DRAW(R.type_mix_by_label_percent[label],rng)
  st := DRAW(R.wlan_fc_subtype_by_label_and_type_percent[label][t],rng)
  assert st in R.encodings.type_to_subtype_allowlists_numeric[t]
  # Radiotap
  f := DRAW(R.channel_frequency_by_label_percent[label],rng)
  if f==5180: (cck,ofdm):=(0,1) else: (cck,ofdm):=DRAW(R.radiotap.observed_24_flags_percent,rng)
  rtlen := DRAW(R.radiotap.radiotap_length_by_label_percent.get(label,R.radiotap.length_mix_percent),rng)
  tsft  := DRAW_BERNOULLI(R.radiotap.tsft_for_length_by_label.get(label,{}).get(rtlen,R.radiotap.tsft_for_length[rtlen]).bern,rng)
  rssi  := DRAW_RSSI(R.rssi_rules[label],rng)
  # MAC fields
  if t in {0,1}: ds:=1 else: ds:=DRAW(R.wlan_fc_ds_rules.data_quota_percent,rng)
  flen := DRAW_FRAME_LEN(R.frame_length_rules,label,rng)
  if t==1: dur:=DRAW(R.duration_rules.control_by_subtype_percent[st],rng)
  else:    dur:=DRAW(R.duration_rules.by_label_percent[label],rng)
  frag:=0; retry:=0; pwrmgt:=0; moredata:=0; prot:=0
  return {"Label":label,"wlan.fc.type":t,"wlan.fc.subtype":st,"wlan.fc.ds":ds,
          "frame.len":flen,"wlan.duration":dur,"radiotap.channel.freq":f,
          "radiotap.channel.flags.cck":cck,"radiotap.channel.flags.ofdm":ofdm,
          "radiotap.length":rtlen,"radiotap.present.tsft":tsft,"radiotap.dbm_antsignal":rssi,
          "wlan.fc.frag":frag,"wlan.fc.retry":retry,"wlan.fc.pwrmgt":pwrmgt,
          "wlan.fc.moredata":moredata,"wlan.fc.protected":prot}

FUNCTION CHECK(R,r) -> bool:
  if r["wlan.fc.subtype"] not in R.encodings.type_to_subtype_allowlists_numeric[r["wlan.fc.type"]]: return False
  if (r["radiotap.channel.freq"]==5180 and not(r["radiotap.channel.flags.ofdm"]==1 and r["radiotap.channel.flags.cck"]==0)) \
     or (r["radiotap.channel.flags.cck"]==1 and r["radiotap.channel.flags.ofdm"]==1): return False
  if (r["wlan.fc.type"] in {0,1} and r["wlan.fc.ds"]!=1) or (r["wlan.fc.type"]==2 and r["wlan.fc.ds"] not in {2,3}): return False
  if r["radiotap.length"]==64 and r["radiotap.present.tsft"]==1: return False
  if r["wlan.fc.type"]==1 and r["wlan.duration"] not in KEYS(R.duration_rules.control_by_subtype_percent[r["wlan.fc.subtype"]]): return False
  if not SUPPORTED_FRAME_LEN(R.frame_length_rules,r["Label"],r["frame.len"]): return False
  return True

PROCEDURE ENFORCE_QUOTAS(R,D):
  D := APPLY_LOCKS(R.protected_joint_policy.locks,D)  # hard locks first
  for label in {0,1,2}:
    D := MATCH_PERCENT(D,"wlan.fc.protected",R.protected_joint_policy.quota_targets.mgmt_percent[label],
                       eligible=MGMT_ROWS(D,label),guards=R.protected_joint_policy.never_touch_when)
    D := MATCH_PERCENT(D,"wlan.fc.protected",R.protected_joint_policy.quota_targets.overall_percent[label],
                       eligible=DATA_ROWS(D,label),guards=R.protected_joint_policy.never_touch_when)
  for F in {"retry","pwrmgt","moredata"}: for label in {0,1,2}:
    D := MATCH_PERCENT(D,"wlan.fc."+F,R.post_flags_quota[F+"_by_label_pct"][label],
                       eligible=ELIGIBLE_FOR_FLAG(D,F,label,R.post_flags_quota.forbid[F]),
                       guards=R.protected_joint_policy.never_touch_when)
  return D
\end{lstlisting}
}

\begin{credits}
\subsubsection{\ackname} This work is supported by the Research Council of Norway through the SFI Norwegian Centre for Cybersecurity in Critical Sectors (NORCICS) project no. 310105.

\subsubsection{\discintname}
The authors have no competing interests to declare that are relevant to the content of this article.
\end{credits}
%
%
%
\bibliographystyle{splncs04}
\bibliography{mybibliography}

@Article{Almorjan_2025,
AUTHOR = {Almorjan, Ashwaq and Basheri, Mohammed and Almasre, Miada},
TITLE = {Large Language Models for Synthetic Dataset Generation of Cybersecurity Indicators of Compromise},
JOURNAL = {Sensors},
VOLUME = {25},
YEAR = {2025},
NUMBER = {9},
ARTICLE-NUMBER = {2825},
PubMedID = {40363263},
ISSN = {1424-8220},
DOI = {10.3390/s25092825}
}

@ARTICLE{AWID3,
  author={Chatzoglou, Efstratios and Kambourakis, Georgios and Kolias, Constantinos},
  journal={IEEE Access}, 
  title={Empirical Evaluation of Attacks Against IEEE 802.11 Enterprise Networks: The AWID3 Dataset}, 
  year={2021},
  volume={9},
  number={},
  pages={34188-34205},
  doi={10.1109/ACCESS.2021.3061609}}

@article{kampourakis2025numeris,
  title={In Numeris Veritas: An Empirical Measurement of Wi-Fi Integration in Industry},
  author={Kampourakis, Vyron and Smiliotopoulos, Christos and Gkioulos, Vasileios and Katsikas, Sokratis},
  journal={arXiv preprint arXiv:2509.16987},
  year={2025}
}

@Misc{80211,
  title = {Part 11: Wireless LAN Medium Access Control (MAC) and Physical Layer (PHY) Specifications.},
  author="IEEE",
  howpublished = "\url{https://standards.ieee.org/ieee/802.11/7028/}"
}

@ARTICLE{zargar2013survey,
  author={Zargar, Saman Taghavi and Joshi, James and Tipper, David},
  journal={IEEE Communications Surveys \& Tutorials}, 
  title={A Survey of Defense Mechanisms Against Distributed Denial of Service (DDoS) Flooding Attacks}, 
  year={2013},
  volume={15},
  number={4},
  pages={2046-2069},
  doi={10.1109/SURV.2013.031413.00127}}

@InProceedings{aminanto2016detecting,
author="Aminanto, Muhamad Erza
and Kim, Kwangjo",
editor="Choi, Dooho
and Guilley, Sylvain",
title="Detecting Impersonation Attack in WiFi Networks Using Deep Learning Approach",
booktitle="Information Security Applications",
year="2017",
publisher="Springer International Publishing",
address="Cham",
pages="136--147",
isbn="978-3-319-56549-1"
}

@article{humby2006data,
  title={Data is the new oil},
  author={Humby, Clive},
  journal={Proc. ANA Sr. Marketer’s Summit. Evanston, IL, USA},
  volume={1},
  pages={1},
  year={2006}
}

@ARTICLE{chatzoglou2022pick,
  author={Chatzoglou, Efstratios and Kambourakis, Georgios and Kolias, Constantinos and Smiliotopoulos, Christos},
  journal={IEEE Access}, 
  title={Pick Quality Over Quantity: Expert Feature Selection and Data Preprocessing for 802.11 Intrusion Detection Systems}, 
  year={2022},
  volume={10},
  number={},
  pages={64761-64784},
  doi={10.1109/ACCESS.2022.3183597}}

@INPROCEEDINGS{Goyal2025,
  author={Goyal, Mandeep and Mahmoud, Qusay H.},
  booktitle={2025 IEEE 15th Annual Computing and Communication Workshop and Conference (CCWC)}, 
  title={An LLM-Based Framework for Synthetic Data Generation}, 
  year={2025},
  volume={},
  number={},
  pages={00340-00346},
  doi={10.1109/CCWC62904.2025.10903878}}

@INPROCEEDINGS{Galadima2024,
  author={Galadima, Haula Sani and Doherty, Cormac and Brennan, Rob},
  booktitle={2024 Cyber Research Conference - Ireland (Cyber-RCI)}, 
  title={Towards LLM-based Synthetic Dataset Generation of Cyber Incident Response Process Logs}, 
  year={2024},
  volume={},
  number={},
  pages={1-4},
  doi={10.1109/Cyber-RCI60769.2024.10939563}}

@ARTICLE{nadas2025,
  author={Nadǎş, Mihai and Dioşan, Laura and Tomescu, Andreea},
  journal={IEEE Access}, 
  title={Synthetic Data Generation Using Large Language Models: Advances in Text and Code}, 
  year={2025},
  volume={13},
  number={},
  pages={134615-134633},
  doi={10.1109/ACCESS.2025.3589503}}

@article{patel2024datadreamer,
  title={Datadreamer: A tool for synthetic data generation and reproducible llm workflows},
  author={Patel, Ajay and Raffel, Colin and Callison-Burch, Chris},
  journal={arXiv preprint arXiv:2402.10379},
  year={2024}
}

@INPROCEEDINGS{swat2016,
  author={Mathur, Aditya P. and Tippenhauer, Nils Ole},
  booktitle={2016 International Workshop on Cyber-physical Systems for Smart Water Networks (CySWater)}, 
  title={SWaT: a water treatment testbed for research and training on ICS security}, 
  year={2016},
  volume={},
  number={},
  pages={31-36},
  doi={10.1109/CySWater.2016.7469060}}

@article{kampDT2025,
  author    = {Kampourakis, Konstantinos E. and
               Gkioulos, Vasileios and
               Kavallieratos, Georgios and
               Lin, Jia-Chun},
  title     = {Digital Twin-Enabled Incident Detection and Response: A Systematic Review of Critical Infrastructures Applications},
  journal   = {International Journal of Information Security},
  year      = {2025},
  volume    = {24},
  number    = {5},
  pages     = {194},
  doi       = {10.1007/s10207-025-01113-0},
  issn      = {1615-5270}
}

@article{kampCR2025,
title = {A step-by-step definition of a reference architecture for cyber ranges},
journal = {Journal of Information Security and Applications},
volume = {88},
pages = {103917},
year = {2025},
issn = {2214-2126},
doi = {https://doi.org/10.1016/j.jisa.2024.103917},
author = {Vyron Kampourakis and Vasileios Gkioulos and Sokratis Katsikas}
}

@ARTICLE{gotham2024,
  author={Sáez-de-Cámara, Xabier and Flores, Jose Luis and Arellano, Cristóbal and Urbieta, Aitor and Zurutuza, Urko},
  journal={IEEE Transactions on Dependable and Secure Computing}, 
  title={Gotham Testbed: A Reproducible IoT Testbed for Security Experiments and Dataset Generation}, 
  year={2024},
  volume={21},
  number={1},
  pages={186-203},
  doi={10.1109/TDSC.2023.3247166}}

@article{kampourakis2025balancing,
  title={Balancing the act? Resampling versus imbalanced data for Wi-Fi IDS},
  author={Kampourakis, Konstantinos E and Chatzoglou, Efstratios and Kambourakis, Georgios and Serpanos, Dimitrios},
  journal={International Journal of Information Security},
  volume={24},
  number={1},
  pages={47},
  year={2025},
  publisher={Springer}
}

@InProceedings{Bl0ck,
author="Chatzoglou, Efstratios
and Kampourakis, Vyron
and Kambourakis, Georgios",
editor="Meyer, Norbert
and Grocholewska-Czury{\l}o, Anna",
title="Bl0ck: Paralyzing 802.11 Connections Through Block Ack Frames",
booktitle="ICT Systems Security and Privacy Protection",
year="2024",
publisher="Springer Nature Switzerland",
address="Cham",
pages="250--264",
isbn="978-3-031-56326-3"
}

@inproceedings{vanhoef-crack-new,
author = {Vanhoef, Mathy and Piessens, Frank},
title = {Release the Kraken: New KRACKs in the 802.11 Standard},
year = {2018},
isbn = {9781450356930},
publisher = {Association for Computing Machinery},
address = {New York, NY, USA},
doi = {10.1145/3243734.3243807},
booktitle = {Proceedings of the 2018 ACM SIGSAC Conference on Computer and Communications Security},
pages = {299–314},
numpages = {16},
location = {Toronto, Canada},
series = {CCS '18}
}

@inproceedings {vanhoef-frag,
author = {Mathy Vanhoef},
title = {Fragment and Forge: Breaking {Wi-Fi} Through Frame Aggregation and Fragmentation},
booktitle = {30th USENIX Security Symposium (USENIX Security 21)},
year = {2021},
isbn = {978-1-939133-24-3},
pages = {161--178},
url = {https://www.usenix.org/conference/usenixsecurity21/presentation/vanhoef},
publisher = {USENIX Association},
month = aug
}

@INPROCEEDINGS{vanhoef-dragonblood,
  author={Vanhoef, Mathy and Ronen, Eyal},
  booktitle={2020 IEEE Symposium on Security and Privacy (SP)}, 
  title={Dragonblood: Analyzing the Dragonfly Handshake of WPA3 and EAP-pwd}, 
  year={2020},
  volume={},
  number={},
  pages={517-533},
  doi={10.1109/SP40000.2020.00031}}
\end{document}